# Temporal pattern of entropy production induced by reactive infiltration instability in natural porous media

Yi Yang[1*], Stefan Bruns[1], Susan L. S. Stipp[1], Henning O. Sørensen[1]

[1] Nano-Science Center, Department of Chemistry, University of Copenhagen, Copenhagen, Denmark

[*] Corresponding author

E-mail: yiyang@nano.ku.dk

# Abstract

The tendency of irreversible processes to generate entropy is the ultimate driving force for the evolution of nature. In engineering, entropy production is often used as a measure of usable energy losses. In this study we show that the analysis of the entropy production patterns can help understand the vastly diversified experimental observations of water-rock interactions in natural porous media. We first present a numerical scheme for the analysis of entropy production in dissolving porous media. Our scheme uses a greyscale digital model of natural chalk obtained by X-ray nanotomography. Greyscale models preserve structural heterogeneities with very high fidelity, which is essential for simulating a system dominated by infiltration instability. We focus on the coupling between two types of entropy production: the percolative entropy generated by dissipating the kinetic energy of fluid flow and the reactive entropy that originates from the consumption of chemical free energy. Their temporal patterns pinpoint three stages of microstructural evolution. We then show that the regional mixing deteriorates infiltration instability by reducing local variations in reactant distribution. In addition, we show that the microstructural evolution can be particularly sensitive to the initially present transport heterogeneities when the global flowrate is small. This dependence on flowrate indicates that the need to resolve the structural features of a porous system is greater when the residence time of the fluid is long.

# Introduction

The production of entropy in all irreversible processes drives the transformations and changes in nature [1, 2]. Spatial and temporal patterns of entropy production can help us understand the diversity and the self-organisation of many complex systems [3, 4]. In engineering, all loses of usable work can be measured in units of entropy generation as a currency [5]. Reactive infiltration instability stems from a positive feedback between the coupling of chemical reaction and mass transfer and induces the development of a wide variety of biological and abiotic flow systems [6-10]. Predicting the development of flow systems in porous media is essential to many energy and environmental applications such as geologic carbon storage [11], oil reservoir stimulation [12], bioremediation [13] and contaminant mobilisation [14]. Characterising the inherent heterogeneities of a porous material constitutes an important step of such predictions because infiltration instability can amplify regional heterogeneities with time indefinitely [15, 16]. It is thus desirable to use non-destructive three-dimensional imaging techniques, like high resolution X-ray tomography [17-20] to obtain greyscale microstructure models of porous media [21, 22]. Based on these models, analysis of entropy production can be carried out with very high fidelity because the geometric complexity is preserved in the greyscale tomographic data [17].

In this study we treat a porous medium as a thermodynamic device that constantly dissipates energy it receives from the environment. The evolution of the device's internal structure is affected by its initial morphology and the energy input. The former dictates the pre-existing heterogeneities in the transport properties of the medium, while the latter describes the different forms of energy that are dissipated by the device. Here we focus on the kinetic energy of fluid flow and the chemical energy of reactive solutes. Fig 1 shows a few examples of a flow system developed by injecting a reactive fluid from the center of a dissolving 2D porous medium. With different fluid inputs, very different morphologies accompanied by different patterns of entropy production can develop from the same initial geometry. Experimentally, such study can be almost impossible to perform because the initial geometries are inevitably destroyed after each experiment. Current manufacturing technology has not yet enabled the replication of natural geologic materials with identical geometry and regional chemical homogeneity on a submicron level. In this study, we analyse numerically the effects of global fluid input on energy dissipation in a natural porous material with a fixed initial geometry. The geometric profile was obtained through a nanometre resolution tomographic scan of a natural chalk sample.

Chalk is a frequent rock in oil and gas reservoirs and a fast reacting model for testing geologic carbon storage in the North Sea. It originates from calcified shields of coccolithophorids that are nanometre-sized and irregular in size, shape and orientation [23, 24]. Characterising the resulting microstructure is difficult because of its geometric complexity that is encoded in the greyscale intensities of a tomography reconstruction. We tried preserving these complexities in the structure that we imaged by avoiding simplification of the microstructure that would usually arise from segmentation of the reconstruction. Instead, we transferred the greyscale intensities to a material density representation that maintains comparability between samples. We then present a full mathematical scheme to use the derived voxel level porosity in simulations of entropy production and show (i) that three stages of microstructural evolution display distinct temporal patterns in entropy generation, (ii) that regional mixing deteriorates reaction frontier instabilities and (iii) that regional heterogeneities become more important with increasing macroscopic flowrate.

Although reactive infiltration instability has been well formulated mathematically by pioneers of geochemical transport modeling [6-9, 15, 16, 25-29], this is the first time a high resolution greyscale nanotomography reconstruction of a natural porous material is incorporated into the irreversibility analysis of an emergent process. This inclusion of realistic geometries is vital because of the sensitivity of the reaction front propagation to any regional heterogeneities [28]. Therefore, by fixing the initial geometry our results not only allow us to outline unambiguously the influence of global energy constraints on morphology evolution, but also make it possible to put the much diversified and sometimes inconsistent experimental observations of reactive percolations into a coherent big picture [30].

## Material and methods

Drill cuttings from the Hod formation (North Sea Basin), identified as sample HC #16, provided air dried chalk particles of ~500 μm diameter that were used for tomographic imaging to generate a model environment close to realistic reservoir conditions. Imaging was performed with the X-ray holotomography setup at the ID22 beamline (29.49 keV) at the European Synchrotron Research Facility in Grenoble, France [31]. The data were reconstructed from 1999 radiographs (360° rotation, 0.5 s exposure) at 100 nm voxel resolution using the holotomography reconstruction method [31] and processed as described in Bruns et al. [32, 33]. The goal of the processing routine was to generate a greyscale volume image where variations in voxel intensity could be related to local material density, i.e. greyscale variations result from partial volume effects and not from signal blur, noise or artefacts. The salient points of the image processing routine were (i) background compensation by Fourier high pass filtering, (ii) noise and artefact removal by de-ringing [34] and iterative non-local means denoising [33], (iii) deconvolution under the assumption of a Gaussian point spread function and (iv) transformation to voxel level porosity by linear interpolation between the average greyscale intensity of chalk and void phase. Since the average intensities of these phases are initially unknown a seven phase Gaussian mixture model was used to identify the most likely intensities for chalk and void phase resulting in a macroscopic porosity value of 0.22 for the 1350×1350×1514 voxel reconstruction. From this reconstruction subvolume samples of 60×60×300 voxels (6×6×30 μm$^3$) were randomly chosen and subjected to a general dissolution reaction A (solid) $\rightleftharpoons$ A (solute).

All physical quantities were normalised to reduce the total number of parameters:

$$q = \frac{Q}{Q_0}, \quad l_n = \frac{l}{L_{ref}}, \quad p = \frac{P}{P_{ref}} \quad \text{and} \quad C = \frac{C_{A,eq} - C_A}{C_{A,eq} - C_{A,inj}}. \quad (1)$$

Where $Q$ is the voxel level volumetric flow rate (m$^3$/s), $l$ is the voxel size (nm) and $P$ is pressure (Pa). $Q_0$, $L_{ref}$ and $P_{ref}$ are their reference values and can be chosen arbitrarily. In this study they were numerically set to the following values for computational effectiveness.

$$Q_0 = \left[ -\frac{(\mu_A^0 + RT)(C_{A,eq} - C_{A,inj})}{k_{A0}^{1/3} \cdot \mu(\varphi_{exp}/k_{exp})} \right]^{1.5}, \quad (2)$$

$$L_{ref} = \sqrt{-\frac{(\mu_A^0 + RT)(C_{A,eq} - C_{A,inj})}{k_{A0} \cdot \mu(\varphi_{exp}/k_{exp})}}, \quad \text{and} \quad (3)$$

$$P_{ref} = -(\mu_A^0 + RT)(C_{A,eq} - C_{A,inj}). \quad (4)$$

Where $\mu_A^0$ is the reference chemical potential of A (kJ/mol), $R$ is the universal gas constant (kJ/mol·K), $T$ is temperature (K), $\mu$ is the viscosity of the fluid (Pa·s), $\varphi_{exp}$ and $k_{exp}$ are the experimentally measured porosity and permeability (m$^2$) of the sample [21], and $k_{A0}$ is the apparent first order rate constant of the dissolution reaction (s$^{-1}$), defined as

$$k_{A0} = \frac{r_{A0}/ssa}{C_0(C_{A,eq} - C_{A,inj})}, \quad (5)$$

where $ssa$ is the voxel level surface ratio (m$^2$/m$^3$) and $C_A$ is the concentration of A at the exit of a voxel (mol/m$^3$). $C_{A,eq}$ and $C_{A,inj}$ are the equilibrium concentration and the concentration of A in the injection fluid, respectively (mol/m$^3$). $C_0$ is the dimensionless concentration at the inlet of a voxel (i.e., corresponding to $C_{A0}$) and $r_{A0}$ is the dissolution rate (mol/m$^2$·s) when $C = C_0$.

The pressure drop across a voxel followed Darcy's law[35] given by

$$\mathbf{q} = -\left[\left(\frac{P_{ref} L_{ref}}{\mu \cdot Q_0}\right)\left(\frac{k_{exp}}{\varphi_{exp}}\right)\right] \cdot \left(l_n \varphi^2\right) \cdot \nabla p, \qquad (6)$$

in which $\varphi$ is the voxel level porosity. The voxel level tortuosity and permeability had been expanded around zero with respect to the empty volume in each voxel. With the first two terms taken the truncation error was expected to decrease with the sixth order of voxel size $l_n$.

For each voxel the continuity equation for an incompressible fluid was applied

$$\nabla \cdot \mathbf{q} = 0. \qquad (7)$$

The macroscopic fluid distribution was computed based on the permeability tensor

$$\mathbf{K} = \mathbf{N_x K_x N_x^T} + \mathbf{N_y K_y N_y^T} + \mathbf{N_z K_z N_z^T}, \qquad (8)$$

where $\mathbf{N_x}$, $\mathbf{N_y}$ and $\mathbf{N_z}$ are nodal matrices along the three Cartesian axis and $\mathbf{K_x}$, $\mathbf{K_y}$ and $\mathbf{K_z}$ are diagonal matrices with voxel level permeabilities on the main diagonal. The pressure field can be calculated using

$$\mathbf{Q_s} = -\mathbf{Kp} \qquad (9)$$

where $\mathbf{Q_s}$ and $\mathbf{p}$ are vectors that describe the global constraint on the system and its pressure field.

In each voxel, the chemical conversion in the subvolume with complete mixing was calculated as

$$-\sum_i q_i C_{0i} + (1 + Da) \cdot qC = 0, \qquad (10)$$

in which *i* indicates the summation over all inlets of reactive fluid and the Damköhler number (Da) is defined as by

$$Da = \left(\frac{r_{A0}}{C_0\left(C_{A,eq} - C_{A,inj}\right)}\right) \cdot \left(\frac{L_{ref}^3}{Q_0}\right) \cdot \eta \cdot l_n^3 \cdot \left(\frac{ssa}{q}\right), \quad (11)$$

where $\eta$ is the portion of a voxel within which reactants and products are completely mixed. The conversion in the non-mixing subvolume was calculated as

$$-C_0 + e^{Da} \cdot C = 0, \text{ with} \quad (12)$$

$$Da = \left(\frac{r_{A0}}{C_0\left(C_{A,eq} - C_{A,inj}\right)}\right) \cdot \left(\frac{L_{ref}^3}{Q_0}\right) \cdot \left(\frac{1-\eta}{3}\right) \cdot l_n^3 \cdot \left(\frac{ssa}{q}\right). \quad (13)$$

The dissolution reaction was assumed to be much slower than establishing the velocity profile and the voxel level porosities were updated according to

$$(\eta \cdot d\varphi)_i = q(C_0 - C) \cdot \left(\frac{C_{A,eq} - C_{A,inj}}{\rho/M}\right) \cdot \left(Q_0 \Big/ L_{ref}^3\right) \cdot \left(\frac{dt}{l_n^3}\right). \quad (14)$$

Percolative entropy generation in each voxel was calculated as

$$\left(\frac{T \cdot \dot{S}_{gen,per}}{Q_0 \cdot P_{ref}}\right) = \frac{q^2}{l_n \varphi^2} \Bigg/ \left[\left(\frac{P_{ref} L_{ref}}{\mu \cdot Q_0}\right)\left(\frac{k_{exp}}{\varphi_{exp}}\right)\right]. \quad (15)$$

The reactive and mixing entropies were not distinguished in this study and were both calculated based on the voxel level Gibbs free energy change:

$$\dot{S}_{gen,rxn} = -\left(\mu_A^0 \Big/ T + R\right)\left(C_{A,eq} - C_{A,inj}\right) Q_0 \cdot q \cdot (C_0 - C). \quad (16)$$

# Results

## Three stages of microstructural evolution that show distinct patterns of entropy generation

Fig 2 shows the initial geometry used in this study and the temporal patterns of entropy generation by different mechanisms. The geometry is a greyscale tomography reconstruction representing the microstructure of a 6 x 6 x 30 μm$^3$ North Sea chalk sample with a volume-averaged porosity of 0.20. The geometry consists of $1.08 \cdot 10^6$ voxels ($100^3$ nm$^3$ each) and the porosity of each voxel is represented by a 32-bit real number. We simulate an injection of reactive fluid from the reactant inlet and sink the effluent at the product outlet. The normalized percolative entropy $S_P$ is an integral of entropy generated over the sample body for overcoming fluid friction and serves as a measure of kinetic energy dissipation. The reactive entropy $S_R$ is calculated based on regional changes of Gibbs free energy and is a measure of chemical energy dissipation.

The temporal patterns of percolative entropy ($S_P$) and reactive entropy ($S_R$) showed three stages of microstructure evolution during a dissolutive percolation: induction, breakthrough and stabilisation (Fig 2b). During induction $S_R$ stayed on a plateau because the residence time was sufficiently long that the fluid reactivity was depleted before reaching the outlet. By monitoring the effluent composition one might naively conclude that the system had reached a steady state. Tracking $S_P$, however, revealed the system dynamics. Percolative entropy generation decreased nonlinearly because mineral dissolution led to the development of a dominant flow path within the complex geometry. This development started from the inlet and advanced gradually downstream, preferentially removing solid materials in the more porous voxels (Fig 3, porosity). This biased removal is characteristic to reactive infiltration instability which amplifies local differences in permeability. The arrival of the reaction front at the outlet (Fig 3, Reactive entropy at dimensionless time $\tau = 50$) marked the start of the breakthrough stage. During breakthrough, a significant increase in macroscopic permeability was accompanied by a shift in the spatial pattern of percolative entropy generation. The newly developed major flow path connected the reactant inlet with the product outlet, thereby channelling the fluid. This minimizes the energy dissipation for overcoming flow resistance. Before this change took place, regions with lower porosity generated more percolative entropy owing to lower permeability (Fig 3, porosity and percolative entropy at $\tau = 50$). After breakthrough, this pattern was inversed by fluid channelling (Fig 3, $\tau = 108$ and onward). Meanwhile, $S_R$ began to decrease as a result of shortened residence time. The last inflection point in $S_P$ marked the end of breakthrough and the beginning of a stabilisation phase. During stabilisation, $S_P$ was dominated by expansion of the major flow path and decreased gradually until the solid was depleted. Reactive entropy generation was limited to regions where sharp gradients of porosity were observed (Fig 3, $\tau = 300$). These regions can be physically interpreted as solid-fluid interfaces, and this spatial pattern change marked a conversion from an advection driven reaction pattern to one dominated by interfacial interactions.

## Regional mixing deteriorates reactive infiltration instability

Fig 4 shows the effect of voxel level mixing on microstructural development. Mixing is the net effect of solute transport. Different transport mechanisms produce different mixing patterns. For example, convection preserves the composition of the bulk flow while diffusion is driven by concentration gradients and tends to "smear out" spatial variations in the solute distribution. Hence, diffusion enhances mixing whereas convection does not. In this study we did not distinguish among transport mechanisms. We instead analysed the net effect of mixing on the chemical conversion of reactants which led to structural morphing. This analysis was done by adjusting the contact pattern between reactant and product on the voxel level. A dimensionless parameter, $\eta$, is introduced for this purpose. $\eta$ is a continuous function that varies between 0 and 1. When $\eta = 0$, there is no mixing between reactant and product when a reaction took place; when $\eta = 1$, a reaction happens only when reactant and product are completely mixed. An analogy can be drawn between $\eta$ and the Péclet number ($Pe$). A $\eta$ close to 1 corresponds to a situation in which $Pe$ is small, *i.e.*, a transport mechanism that enhances mixing (e.g., molecular diffusion) is predominant. Similarly, a $\eta$ close to 0 corresponds to a large $Pe$. The caveat of adopting $\eta$ is that one cannot predict precisely the system behaviour – instead, $\eta$ allows "bracketing" the possible outcomes of a real system by two extreme cases ($\eta = 0$ and 1). Assigning a single value of $\eta$ to all voxels in a dataset also ignores regional heterogeneity in the strength of molecular transport, which might not be appropriate when the scale of the system in question is large. Nevertheless, we consider $\eta$ is a general descriptor of transport phenomena in this study because it is not limited to any specific combination of transport mechanisms and because of the small scale concerned.

Mixing protracted the development of the major flow path and delayed the occurrence of the breakthrough (Fig 4a). This delay stemmed from the fact that mixing counteracted the instability that drove the morphing of the system. The evolution of the microstructure, especially the development of the major flow path, was induced by reactive infiltration instability. This instability amplified local heterogeneities in material properties by providing a positive feedback between mass transfer and mineral dissolution. The net observable effect was autocatalytic – both flow and reaction enhanced themselves regionally through mutual coupling. Mixing introduced a negative feedback to this chain of coupling by reducing the reactant concentration and therefore the rate of the dissolution reaction. This negative feedback can only be observed when the reaction rate increases monotonically with chemical affinity. This prerequisite is met by a wide range of geochemical reactions that follow a transition state theory (TST) based rate law.

The effect of mixing on $S_P$ escalated as the main flow path developed towards the outlet, reaching a maximum during the breakthrough stage and diminished rapidly during stabilisation. This variation was attributed to the sensitivity of the percolation entropy generation to the geometric complexity of a porous medium before the development of a main flow path. This flow path advanced faster without mixing, hence lower $S_P$ with smaller $\eta$. For example, at a volume averaged porosity of 0.53, a 43% difference between $\eta = 0$ and 1 was observed for mechanical energy dissipation, suggesting that more energy was needed to drive the same amount of fluid through an evolving microstructure when regional mixing was increased. After breakthrough, the pressure drop was determined predominantly by the permeability of the major flow path, and $S_P$ became insensitive to the mixing status within the solid residual.

Figs 4b and 4c show cross sections (X = 30 μm) of the geometry and the corresponding spatial patterns of reactive entropy generation in two limiting cases of mixing. Regions of interest (ROIs) were highlighted with rectangular boxes. The temporal patterns of $S_R$ suggest that with decreased mixing the reactive fluid was more "penetrating" and the reaction front reached the outlet with less solid dissolved (upper right inset in Fig 4a). This effect was manifested in Fig 4b – given the same overall porosity, a stronger mixing ($\eta = 1$) resulted in a more thorough dissolution of the upstream materials (ROI1) but less developed pore structures downstream (ROIs 2-4). As a consequence, mixing led to a longer residence time while leaving fewer voxels partially filled with dissolvable solid. These two aspects exerted opposite influences on reactive entropy generation. Longer residence times enhanced the chemical conversion which in turn increased $S_R$. Meanwhile, a decrease in the number of intermediate grey voxels can be physically interpreted as a smoothening of the material's surface, which leads to a drop in reactive surface area and hence, $S_R$. These counteracting factors resulted in a complex temporal pattern of entropy generation. Before the breakthrough, the residence time dominated the chemical conversion, and a system with better mixing dissipated more chemical energy. After the breakthrough, the residence times of all systems quickly converged to a single value that was determined by the major flow path. Therefore, the surface roughness developed during earlier stages is the predominant factor and a system with lesser mixing would end up producing more reactive entropy than a well-mixed system. Fig 4a shows this transition between these cases: the apparent insensitivity of $S_R$ to mixing during the breakthrough stage was actually a superposition of two contradicting effects and during the stabilisation stage $S_R$ was greater for systems with higher surface roughness. Fig 4c shows that while the overall $S_R$ was similar for both systems at $\varphi = 0.5$, lesser mixing led to weaker but more dispersed entropy hotspots (ROIs 5-7). In contrast, when $\eta = 1$ the system showed sharp and bright entropy hotspots along the interfaces upstream, but far less downstream. In general, more mechanical energy is needed to drive the evolution of a porous medium with greater regional mixing towards complete dissolution.

## Greater flowrate reduces impact of initial heterogeneities

In this study porous media are considered thermodynamic devices that dissipate energy they receive from the environment. It is thus of interest to investigate the effects of a global constraint, i.e. the total amount of energy that the media receive, on the microstructural evolution. We used the macroscopic flowrate as a measure of this global constraint because it reflects both the energy requirement for overcoming the resistance to the fluid flow and the amount of chemical reactant input given the same solvent composition. Given the size of a simulation domain, a greater flowrate indicates a shorter residence time for the medium to "digest" fluid reactivity. Fig 5 shows the impact of the macroscopic flowrate $Q$ (dimensionless) on the temporal and spatial patterns of entropy generation. Both percolative and reactive entropies scale with the flowrate because with greater throughput the medium received more mechanical and chemical energy. $S_P$ increased with $Q$ because more energy is required to drive a greater volumetric flow through the same microstructure. The temporal pattern of $S_P$ was most sensitive to the geometric complexity when $Q$ was low, whilst with a greater $Q$ the system resembles a homogeneous medium for which the various stages of structural evolution were not distinct. For example, in Fig 5a the last inflection point of $S_P$ can be easily identified for $Q = 0.1$ (at $\varphi = 0.53$), but not for $Q = 10$ (at a $\varphi$ less than 0.4). A close examination of the geometric cross sections (Fig 5b) showed that a high fluid throughput resulted in greater surface roughness (manifested by a higher number of partially filled voxels). In contrast, the cross section of $Q = 0.1$ showed that the original heterogeneities have been amplified, leaving sharp interfaces and a more thoroughly dissolved upstream geometry. Fig 5c shows that the channelling effect was less observable for $Q = 10$, in which many less porous regions were still percolative entropy hotspots (ROI 1).

The impact of the global flowrate is also evident in the patterns of reactive entropy production. Fig 5d shows that the temporal pattern of $S_R$ can also be divided into three stages. In the first stage $S_R$ was limited by the fluid reactivity, which, given an identical reactant concentration at the inlet, scaled with the fluid throughput. The second stage showed a gradual decrease of $S_R$ as the major flow path expanded and the fluid residence time shortened. The last stage featured a quick drop of $S_R$, which stemmed from the depletion of solid reactivity as the macroscopic porosity approached unity (complete dissolution). When less reactant was available solid reactivity depletion was delayed, and vice versa. When $Q = 10$ the third stage of $S_R$ started before the macroscopic porosity reached 0.8, while for $Q = 2.0$ this change took place at approximately $\varphi = 0.9$. A visualisation of regional Damköhler numbers (Fig 5e, Eqs 11 & 13) shows that a greater fluid throughput drove the reactants into less permeable regions and thus making more space available for chemical reactions (ROI 2). However, the chemical conversions in these voxels were low because of the short residence time, leaving many partially dissolved voxels behind. The spatial pattern of $S_R$ (Fig 5f) shows clearly that the reaction was more convection-driven and dispersive when $Q = 10$ and was more interface-dominant when $Q = 0.1$.

The effects of the global constraint on microstructural evolution have two important practical implications. First, the amplification of regional heterogeneity during structural morphing is $Q$ dependent. The distinguishability of different stages in the temporal entropy generation is a measure of system sensitivity to the initial geometry. Fig 5 shows that this distinguishability depends heavily on the global constraint, which implies that the reactive infiltration instability is important in big systems where the fluid input can be considered small and regional. Meanwhile, the prediction of system evolution with a large $Q$ requires lower resolving capability of the initial geometry. Second, percolative entropy generation is closely related to the macroscopic permeability of a porous medium. When the evolution of a system needs to be accounted for, a representative elementary volume may have to be defined according not only to the nature of the porous material but also to the operation environment that dictates the global constraint.

## Discussion

Irreversibility analysis based on grey scale tomography reconstructions provides unprecedented insights into the structural development of a natural microfluidic system. Assuming chemical homogeneity, the evolution of a system is controlled by only four dimensionless parameters: voxel level porosity ($\varphi$), dimensionless voxel size ($l_n$), extent of regional mixing ($\eta$) and the global constraint (measured by $Q$). Whereas $\varphi$ and $l_n$ are dictated by the tomographic data, $\eta$ and $Q$ can vary according to the nature of the porous material and the operational conditions of a reactive percolation. This variability can help us decouple the effects of geometric complexity from those of other factors by conducting numerical experiments under various conditions on a totally fixed geometry. Moreover, it also allows us to revisit many previous experimental observations and put the results into a more coherent bigger picture.

A defining moment during the instability induced microstructural evolution is the swapping spatial pattern of percolative entropy generation during breakthrough (*e.g.*, Fig 3, $S_P$ pattern at $\tau = 50$ vs $\tau = 108$). It leads to distinct system behaviours that had been interpreted differently and given various names. Before the breakthrough, the major flow path develops from the reactive fluid inlet and slowly advances downstream. Because of the system's tendency to amplify regional heterogeneity, this path development is usually accompanied by the preferential removal of the more permeable or more reactive materials. For example, microcrystalline and concomitant particles have been observed to dissolve preferentially during the early stage of acidic percolations of limestone [36, 37]. The sharpening of solid-liquid interfaces after breakthrough, most clearly shown in the spatial patterns of reactive entropy generation (e.g., Fig 3, $\tau = 300$ and Fig 5f), has also been observed tomographically by registering boundary geometry [30, 38]. This interface-focused reaction pattern accompanying the expansion of the major flow path has been referred to as sparitic dissolution [36], interface smoothing [37], surface/transport control [39] or heterogeneous dissolution [40]. In contrast, the convective and dispersive pattern before breakthrough has been referred to as reaction control or uniform/homogeneous dissolution [30]. The homogeneous/heterogeneous categorisation is particularly informative because it not only describes the spatial patterns of structural evolution but also implies that as $Q$ approaches infinity the effect of inherent geometric heterogeneity on morphology development will vanish (Fig 5a), i.e. a porous medium would evolve as if homogeneous. It is worth emphasising that homogeneous and heterogeneous dissolution patterns can coexist and are separated spatially by the advancing frontier of the developing major flow path (e.g., Fig 3, $S_R$ pattern at $\tau =108$ and Fig 5f, $Q = 10$). The breakthrough event, that is marked by the last inflection point in the temporal $S_P$ pattern, has been documented experimentally as a transitional regime [40]. Physically the breakthrough means the connection of the percolation inlet and outlet by the newly developed major flow path (Fig 3). As a consequence, the macroscopic permeability increases sharply with a minor change in porosity [40]; the pore connectivity increases[36, 37] as small pores get interconnected through the major flow path; the volume averaged tortuosity decreases [40] because of the shortened residence time; and both the effective porosity [41] – defined as the porosity in the convective subvolume of a sample – and the effective hydraulic radius increase [40] because of the channelling of the fluid into the major flow path.

The temporal pattern of $S_R$ helps reconcile inconsistencies between experimental observations based on solution chemistry analysis. The value of $S_R$ reflects the overall dissipation of chemical free energy entering the system and can thus be related to the outlet concentrations of reactants (e.g., pH) or products (e.g., metal cations). Percolation experiments may cover different stages of the $S_R$ evolution and observe inconsistent trends in outlet concentration and – as a derivative – reactive surface area. This apparent contradiction can be best interpreted by the relative reactivity between the fluid and the solid (c.f., Fig 1). Fig 6 shows the effects of $l_n$, used as a measure of system size, on entropy generation patterns. A small $l_n$ corresponds to a porous medium with the same geometric complexity but smaller in scale (and thus containing less solid material). If percolation is initially limited by fluid reactivity (Fig 6b, $l_n = 0.5, 1.0, 5.0$ and $10.0$) the effluent will be saturated with minerals dissolved from the porous medium [42]. This quasi steady state corresponds to the initial plateau in reactive entropy generation. As the pore structure develops, the outlet concentrations of dissolution products will decrease exponentially as a result of shortened residence time [36, 39, 40] and eventually drop quickly to zero as the reaction becomes limited by solid reactivity (e.g., as $\varphi$ approaches one when the microstructure will be dissolved completely). However, if percolation is initially limited by solid reactivity – as very often seen in aluminosilicate dissolutions [43, 44] – it is possible for the reactive surface area to increase over time as more pores open up that enable interfacial contact (Fig 6b, initial $S_R$ increase for $l_n = 0.1$). If a percolation starts from the stabilisation stage with pre-existing major flow paths (e.g., fractures in rocks), mineralogical heterogeneity may result in preferential removal of the more reactive materials and thus produce a surface roughening (in contrast to surface smoothening when assuming chemical homogeneity) that increases the contact between fluid and the slow-dissolving minerals[45]. Natural porous media usually consist of multiple mineral phases that differ in reactivity. It is thus expected that a single dissolutive percolation may display different trends for the concentrations of dissolution products from minerals with different reactivities. For example, Noiriel et al. observed a decrease in micrite (fast) dissolution product and an increase in sparite (slow) dissolution product over time in the same percolation experiment [38].

Fig 6 also suggests that the effects of the global constraint (measured by $Q$) and that of the system size (measured by $l_n$) are equivalent. A greater $l_n$ corresponds to a smaller $Q$ given the same chemical free energy input because it provides a longer residence time for the reactive fluid. A small system ($l_n = 0.1$, Fig 6c and 6d) dissolves "homogeneously" while a big system ($l_n = 10$) demonstrates a spatial separation between homogeneous and heterogeneous regimes by the dissolution frontier. Experimentally the effect of $Q$ can be measured by a macroscopic Damköhler number. This number can be changed through tuning the reaction rate (e.g., by manipulating the partial pressure of $CO_2$) [39] or by varying residence time (e.g., by adjusting the flow rate) [37]. However, this number reflects the relative amount of the two forms of energy a system can dissipate (mechanical vs chemical) and should not be considered identical to regional Damköhler number (Figs 5e) which controls local reactant conversion.

# Conclusions

We presented a mathematical scheme to analyse the patterns of entropy production in a dissolving natural porous medium based on numerical simulations with a fixed initial microstructure. This approach uses a greyscale digital model of natural chalk obtained with X-ray nanotomography. Greyscale models allow the preservation of structural heterogeneities with very high fidelity. This is especially important for simulating systems dominated by infiltration instability because infiltration instability amplifies regional differences in material density. We studied two types of entropy production: the percolative entropy generated by dissipating the kinetic energy of fluid flow and the reactive entropy originated from the consumption of chemical free energy. By analysing these temporal patterns, we identified three distinct stages of microstructural evolution in a dissolving porous medium: induction, breakthrough and stabilisation. We also found that regional mixing dampens infiltration instability by reducing local variations in reactant distribution. In addition, we showed that the microstructural evolution can be particularly sensitive to any initial transport heterogeneities when the global flowrate is small. This dependence on flowrate indicates that the need to resolve the structural features is greater when the residence time of the fluid is long, and vice versa. We concluded the discussion by presenting a few examples of how patterns of entropy production can help understand and unify the vastly diversified experimental observations of water-rock interaction in natural porous media.

## Acknowledgments

Parts of this research were carried out at the ID22 beamline at ESRF (European Synchrotron Research Facility, France). We thank H. Suhonen at for technical support. We are grateful to F. Engstrøm from Maersk Oil and Gas A/S for providing the sample. Funding for this project was provided by the Innovation Fund Denmark, through the CINEMA project, the Innovation Fund Denmark and Maersk Oil and Gas A/S, through the P3 project as well as the European Commission, Horizon 2020 Research and Innovation Programme under the Marie Sklodowska-Curie Grant Agreement No 653241. We thank the Danish Council for Independent Research for support for synchrotron beamtime through Danscatt.

**Fig 1. Example morphologies of developing flow systems dominated by infiltration instability.** A dissolutive fluid is injected from the centre of the 2D domain consisting of 4 million reacting units. The colour corresponds to the apparent solid dissolution rate (yellow means more dissolution). (a) vs. (b): fluid reactivity-limited vs. solid reactivity-limited system. (c) vs. (d): small vs. large injection rate.

**Fig 2. Temporal pattern of percolative and reactive entropy production based on a 32-bit greyscale tomography reconstruction of natural chalk.** (a) Perspective view of the simulation domain with an initial porosity of 0.20. (b) Temporal patterns of entropy generation divide the microstructural evolution into three stages: induction, breakthrough and stabilisation. The volume averaged porosity is used as a measure of overall reaction progress.

**Fig 3. Spatial patterns of entropy production during different stages of microstructural evolution.** The rectangular and square images are cross sections of the corresponding quantities at the middle of the radial and the axial directions. The pattern of percolative entropy generation was inversed after the breakthrough because of fluid channelling. The spatial patterns of reactive entropy visualise the dissolving regions in the pore structure.

**Fig 4. Impact of voxel level mixing on entropy production.** $\eta = 1$ corresponds to complete mixing and $\eta = 0$ no mixing between reactant and product. (a) Mixing effect on the temporal patterns of entropy generation. Percolative entropy was affected monotonically by mixing while reactive entropy appeared to be affected by two counteracting effects. (b) Cross sections of porosity distribution. Sharpened solid-liquid interface is evident near the fluid inlet. (c) Distribution of reactive entropy production shows highly localised dissolving regions. Brighter means more dissolution. Mixing leads to a longer residence time and higher surface smoothness. (b) and (c) were taken when the macroscopic porosity was 0.5.

**Fig 5. Effects of macroscopic flow rate ($Q$) on microstructural evolution.** (a) Temporal patterns of percolative entropy generation. (b) Cross sections of porosity distribution at different $Q$. (c) Spatial patterns of percolative entropy generation with region of interest (ROI) 1 showing entropy hotspots in a low porosity subvolume when the throughput was high. (d) Temporal patterns of reactive entropy generation. (e) Regional Damköhler number with ROI 2 showing distinct spatial patterns of $Da$ in a low porosity subvolume. (f) Spatial patterns of reactive entropy generation. All cross sections were taken when macroscopic porosity was 0.5 with the upper image showing $Q = 10$ and the lower showing $Q = 0.1$.

**Fig 6. Effect of system size on entropy generation.** System size is changed by manipulating the dimensionless voxel size ($l_n$) while keeping the geometry constant. A smaller $l_n$ corresponds to a system with less solid material and, given the same fluid throughput, a greater fluid residence time. (a) and (b): temporal patterns of entropy production. (c) and (d): cross sections of porosity and entropy generation when macroscopic porosity equals 0.5. A smaller system dissolves more "homogeneously", concurring with experimental observations.

**Figure 1**

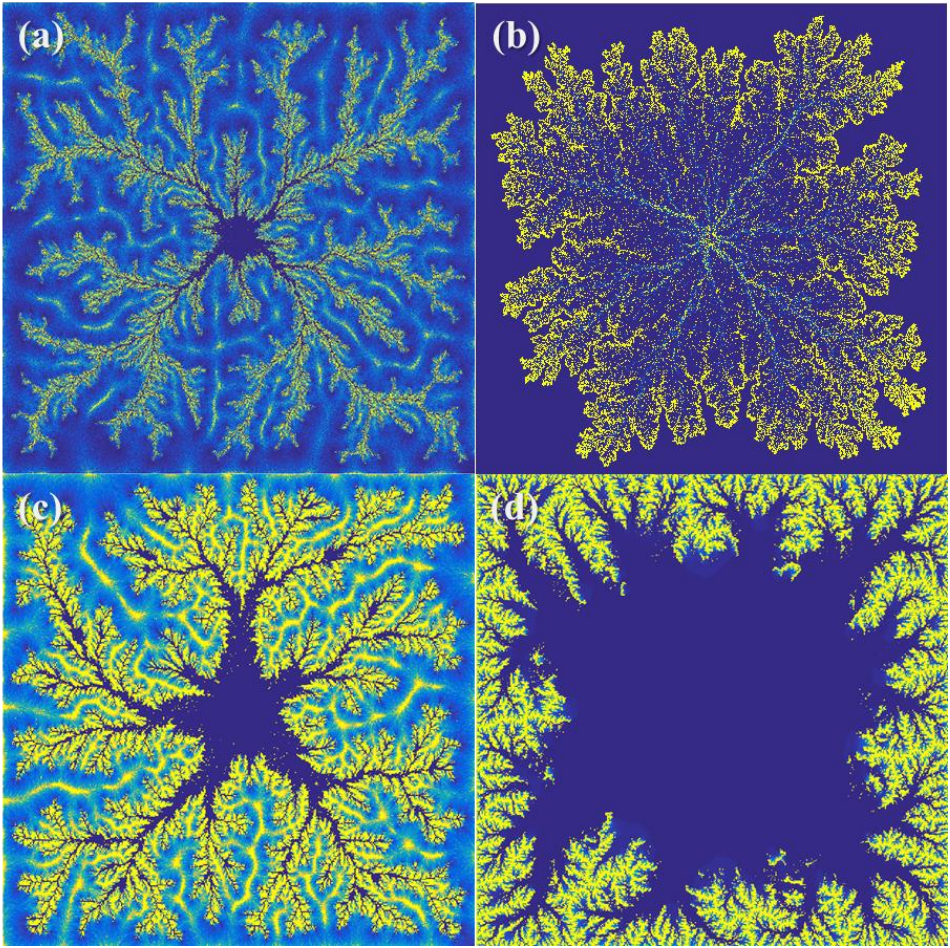

**Figure 2**

**(a)**

$\varphi = 0.20$
**Number of voxels: 1,080,000**
**Data type: 32-bit Real**

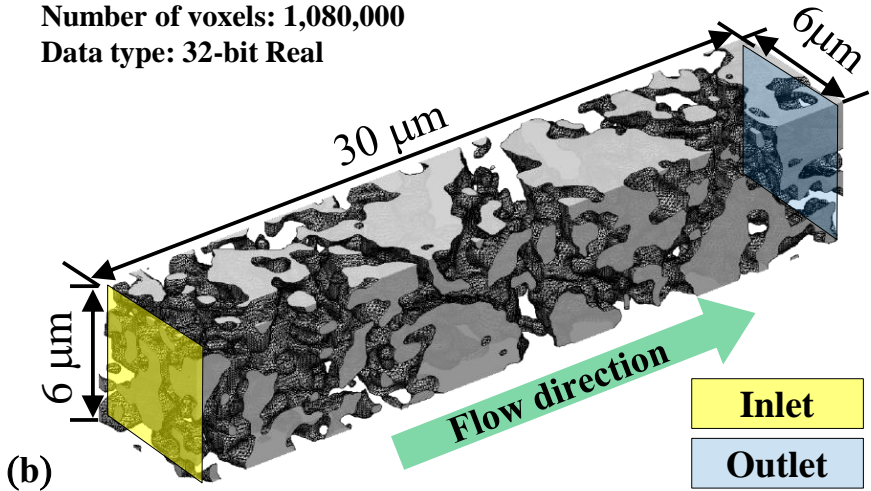

**(b)**

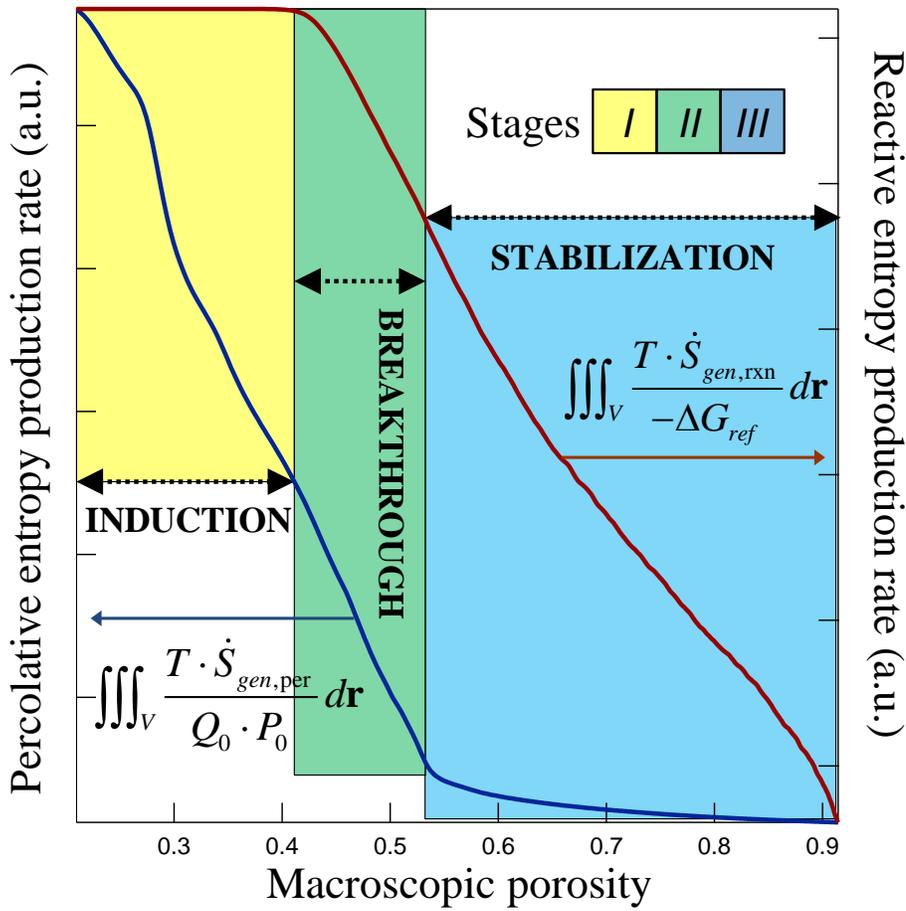

**Figure 3**

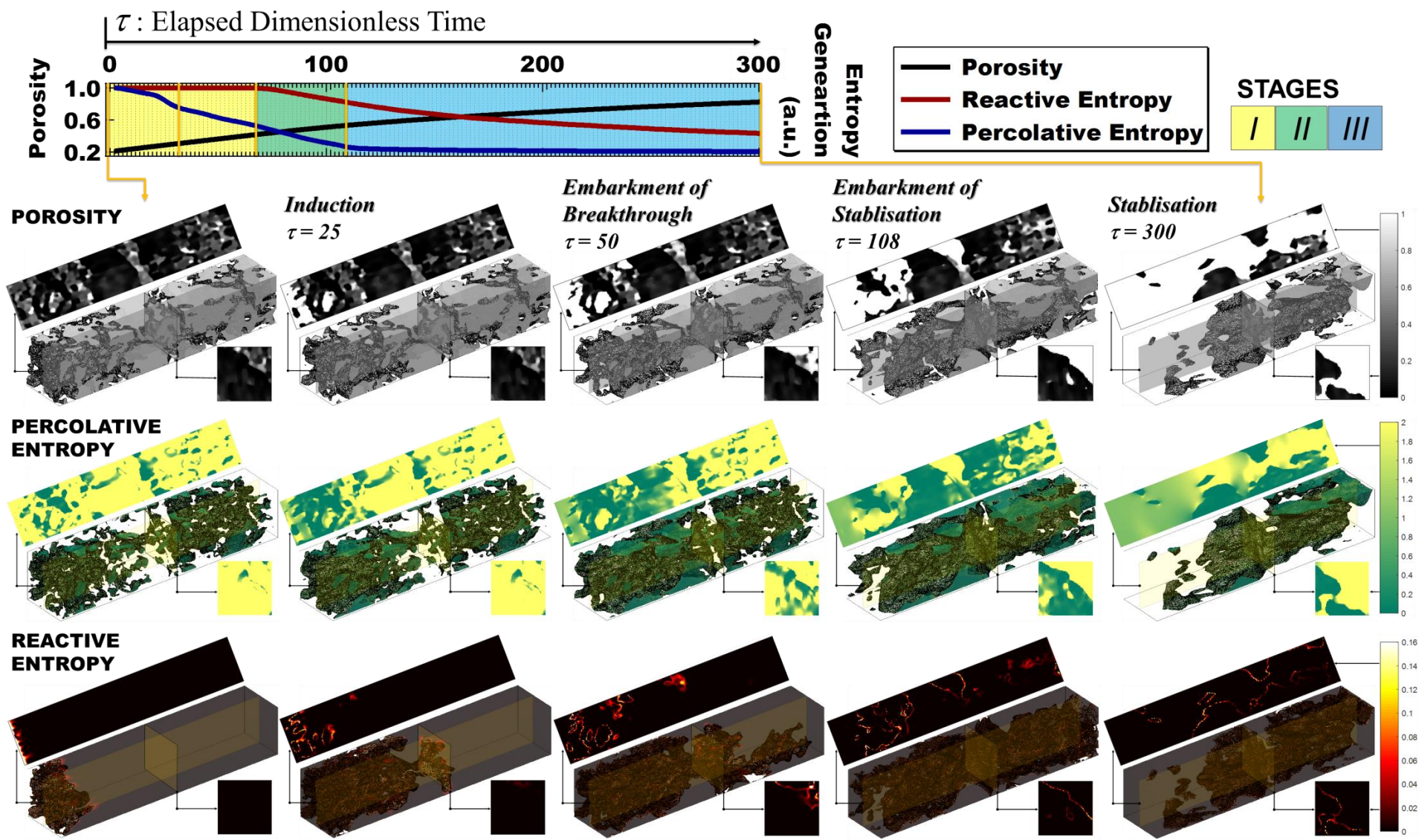

**Figure 4**

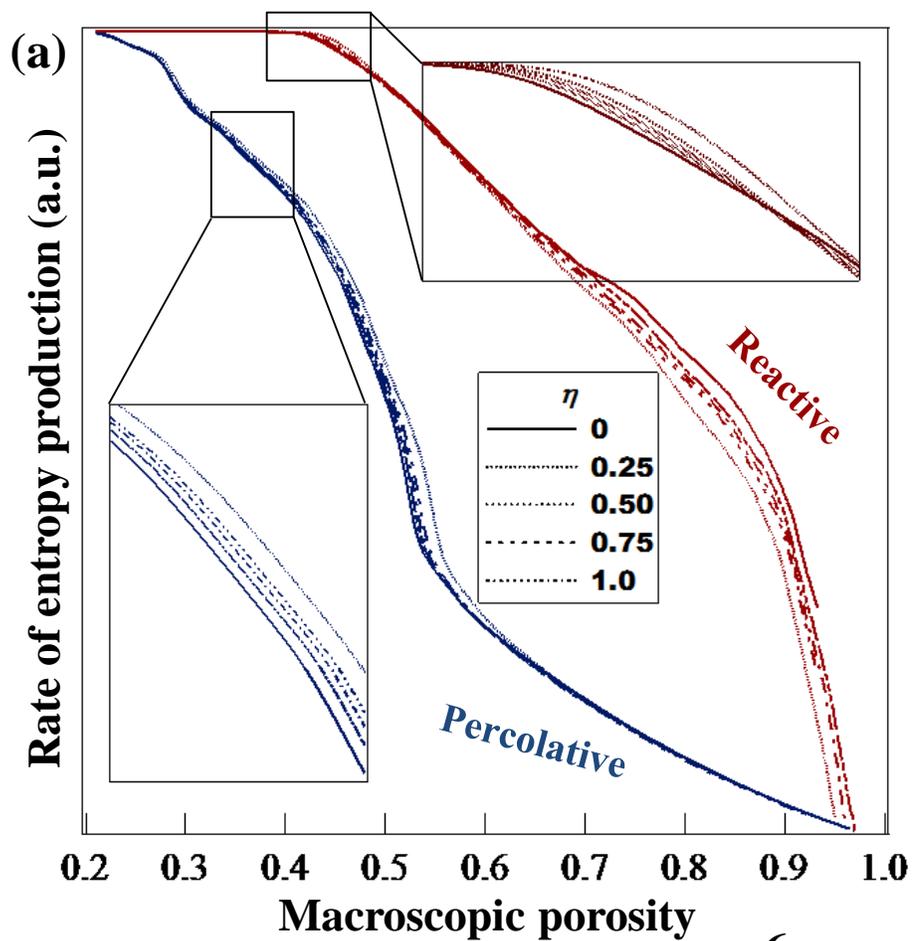

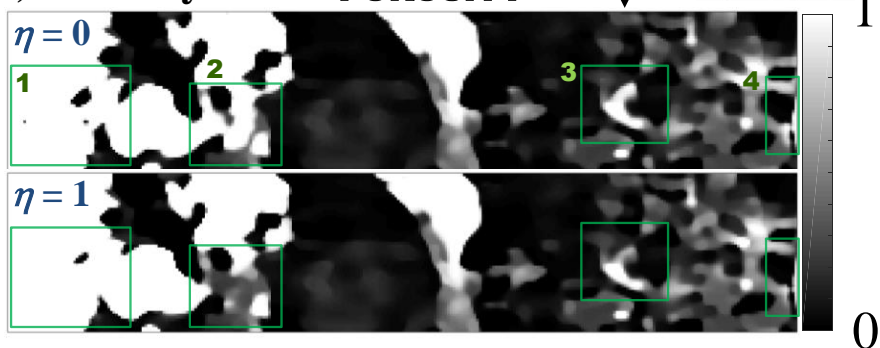

(b) Porosity distribution

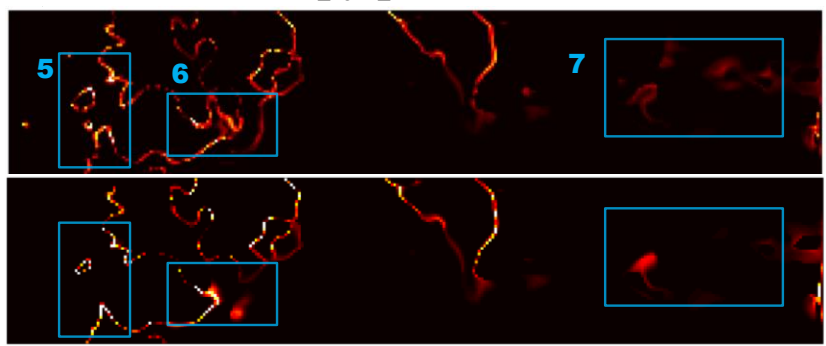

(c) Reactive entropy production

**Figure 5**

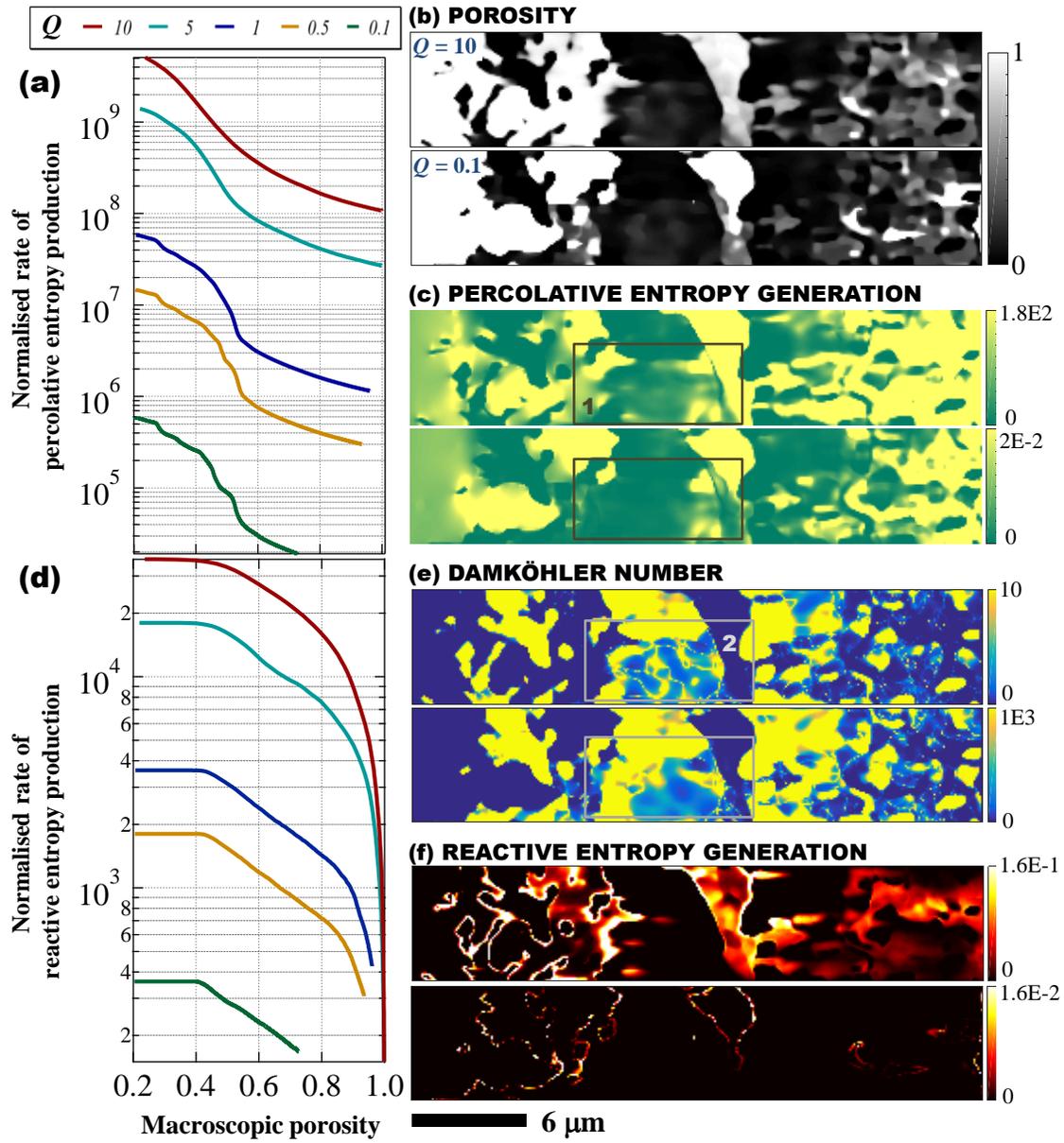

**Figure 6**

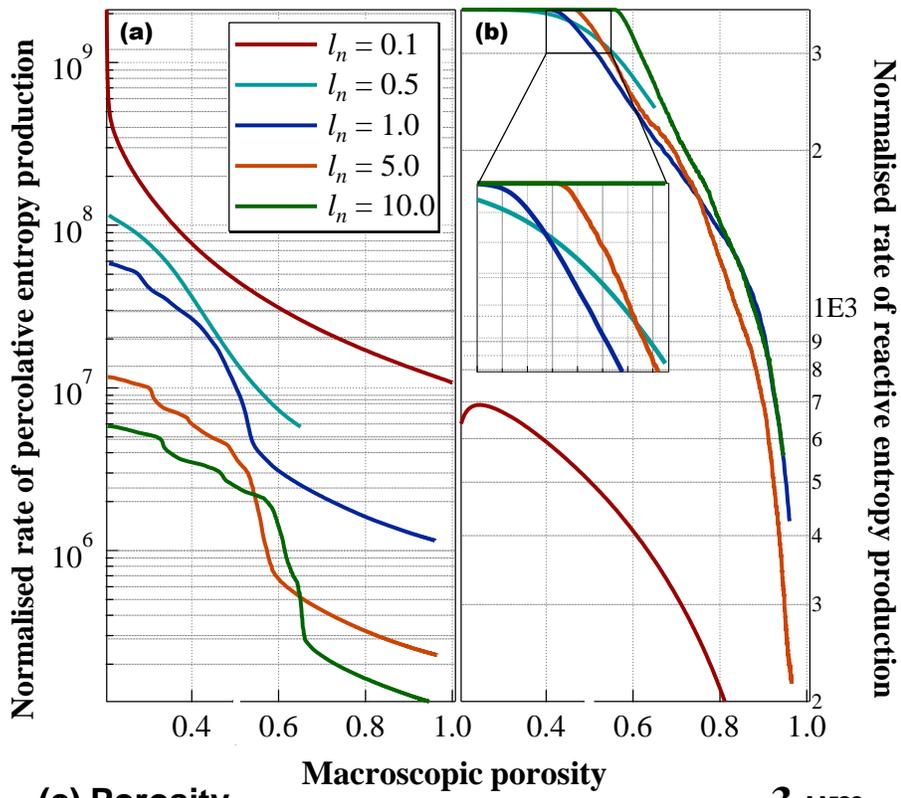

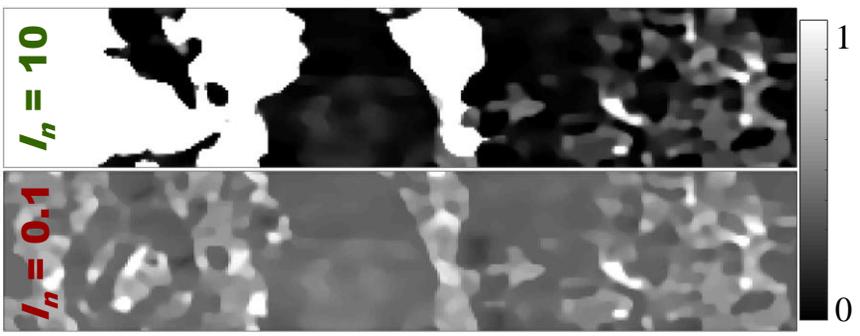

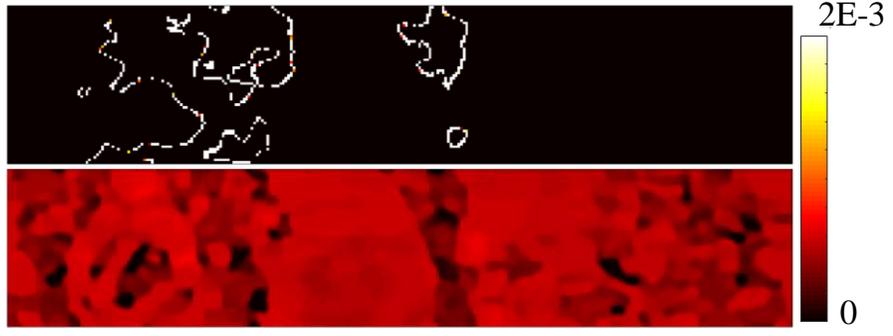